\documentclass[aip,jap,showpacs,showkeys,
superscriptaddress, floatfix,
twocolumn,
reprint]{revtex4}
\usepackage{amsmath}
\usepackage{amsfonts}
\usepackage{amssymb}
\usepackage{graphics}
\usepackage{float}
\usepackage{graphicx}
\usepackage{epstopdf}
\usepackage[compact]{titlesec}
\usepackage{natbib}
\providecommand{\U}[1]{\protect\rule{.1in}{.1in}}
\titlespacing{\section}{0pt}{2ex}{1ex}
\titlespacing{\subsection}{0pt}{0.5ex}{0ex}
\begin{document}
\title{Design principles for HgTe based Topological Insulator Devices}
\author{Parijat Sengupta}
\email{psengupta@purdue.edu}
\author{Tillmann Kubis}
\author{Yaohua Tan}
\author{Michael Povolotskyi}
\author{Gerhard Klimeck}
\affiliation{Dept of Electrical and Computer Engineering, Purdue University, West
Lafayette, IN, 47907}

\begin{abstract}
The topological insulator properties of CdTe/HgTe/CdTe quantum
wells are theoretically studied. The CdTe/HgTe/CdTe quantum
well behaves as a topological insulator beyond a critical well width dimension. It is shown that if the barrier(CdTe) and
well-region(HgTe) are altered by replacing them with the alloy Cd$_{x}%
$Hg$_{1-x}$Te of various stoichiometries, the critical width can be
changed.The critical quantum well width is shown to depend on
temperature, applied stress, growth directions and external electric fields.
Based on these results, a novel device concept is proposed that allows to
switch between a normal semiconducting and topological insulator state through
application of moderate external electric fields.

\end{abstract}
\maketitle

\section{Introduction}

\label{sec:intro} An insulator is conventionally defined as a material that
does not conduct electricity. In most insulators the lack of electric
conduction is explained using its bandstructure properties.The band theory predicts
that an insulator has an energy gap separating the conduction and valence
bands. As a result of this finite band-gap there are no electronic
states to support the flow of current. Recently, materials that have an energy
gap in bulk but possess gapless states bound to the sample surface or edge
have been theoretically predicted and experimentally observed.~\cite{hasanrmp}
These states, in a time reversal invariant system are protected against
perturbation and nonmagnetic
disorder.~\cite{fuprb07,roy09,kaneqsh05,murakami07} Materials that support
such states are known as topological insulators (TI). Examples of
materials with such properties include Bi$_{2}$Te$_{3}$, Bi$_{2}$Se$_{3}$,
Bi$_{x}$Sb$_{1-x}$ alloys, and CdTe/HgTe/CdTe quantum wells. Bi$_{2}$Te$_{3}%
$,Bi$_{2}$Se$_{3}$, and Bi$_{x}$Sb$_{1-x}$ belong to the class of 3-D
topological insulators (3D-TI) and host bound states on their
surface.~\cite{zhang09,xia09obsv,chen09exp} CdTe-HgTe-CdTe quantum wells,
which were the first predicted TIs are 2-D topological insulators (2-D TI).
Unlike their 3D counterpart, they possess bound states at the edge of the
quantum well.~\cite{könig07,bernevig06,roth09hgte} These conducting surface
and edge states develop at the boundary between two insulators, where one is
normal (NI) and the other of inverted band ordering. The surface states, which
are subject to the details of the band properties of each involved material
can mutually influence each other. It is therefore essential to theoretically
study the surface states under various conditions.

This work proposes ways that can efficiently invert the band
profile of a CdTe/HgTe/CdTe heterostructure and create bound edge states.
Specifically, the transition from an NI to a TI through external adiabatic
parameters, adjustable lattice constants, or modulation of the electron-hole
band coupling is the underlying theme. The paper is organized as
follows: In Section~\ref{sec:materials}, details of the CdTe/HgTe/CdTe
structure and method to compute the energy dispersion are described.
Section~\ref{sec:results} discusses various inverted band structures, device
and material conditions that can lead to topologically protected conducting
surface states. The concept of creating a switch from a topological insulator
is developed here. The key results discussed in the paper are summarized in
Section~\ref{sec:conc}.

\section{Materials and Methods\label{sec:materials}}

\begin{figure}[h]
\includegraphics[scale=0.85]{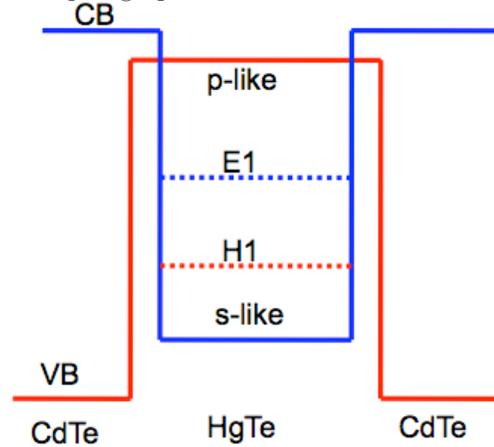}
\caption{Sketch of a CdTe/HgTe/CdTe quantum well heterostructure. The lowest conduction band (CB) state is labeled
with E1 and the highest valence band (VB) state with H1.}
\label{fig1}
\end{figure}

An HgTe quantum well flanked by CdTe barriers has been shown to have edge
states with topological insulator properties.~\cite{buttiker2009edge} TI
behaviour is possible because CdTe is a normal insulator and is placed in
contact with an inverted insulator HgTe. A representative sketch of the device
is shown in Fig.\ref{fig1}. CdTe is a wide band gap semiconductor
($E_{g}=1.606~\mathrm{eV}$) with positive energy gap (NI) and a small lattice
mismatch of 0.5\% with HgTe. CdTe, because of similar lattice constants is chosen as the barrier for the
HgTe well region though in principle any normal ordered material would suffice. The normal valence and conduction band are reversed in their energetic order in HgTe as indicated in Fig.~\ref{fig1} and explained in the next paragraph. 

\begin{figure}[h]
\includegraphics{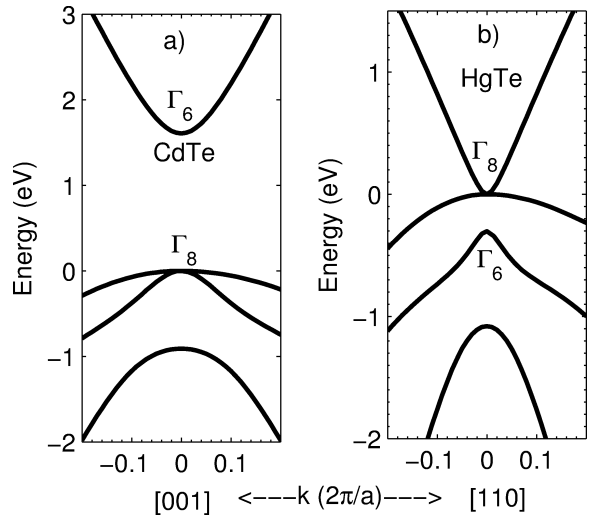}
\caption{Bulk band structure of CdTe (a) and HgTe(b). The ordering of the conduction and valence bands near the band gap at the
$\Gamma$ point in HgTe (Fig.~\ref{fig2}b) is opposite to the one in CdTe (Fig.~\ref{fig2}a). In HgTe, the hole state $\Gamma_{8}$ is above the
electron state $\Gamma_{6}$.}
\label{fig2}
\end{figure}

The inversion of bands for the CdTe/HgTe/CdTe heterostructure is achieved
through the HgTe component. Both CdTe and HgTe belong to the zinc blende (ZB)
structure with T$_{d}$ point group symmetry. The highest valence and lowest
conduction band is made up of \textit{p} and \textit{s} orbitals respectively.
A \textit{normal} band order at $\Gamma$ has lowest conduction band ($j=1/2$)
with $\Gamma_{6}$ symmetry above the top of the valence bands ($j=3/2$) with
$\Gamma_{8}$ symmetry. The $\Gamma_{6}$ state has \textit{s}-type symmetry and
the $\Gamma_{8}$ state has \textit{p}-type symmetry. In a normal ordered
material $\Gamma_{6}$ state is energetically higher than the $\Gamma_{8}$
state. This order is reversed in bulk HgTe at the $\Gamma$ point due to the
high spin-orbit coupling and a significant Darwin term
contribution.~\cite{cade1985self} The strong spin orbit coupling pushes the
valence bands upwards while the Darwin term shifts the \textit{s}-type
conduction band down. The Darwin term can only influence the \textit{s}-type
bands.~\cite{im1982} The combined effect of spin orbit coupling and Darwin
term yields an inverted band order at the $\Gamma$point which flips the order
of the high-symmetry $\Gamma_{6}$ and $\Gamma_{8}$ points for
HgTe.~\cite{molenkampprivate}The energy gap at $\Gamma$ which is defined as

\begin{equation}
E_{g}=E(\Gamma_{6})-E(\Gamma_{8}),
\label{eq:fullham}
\end{equation}
therefore turns out to be negative for HgTe. The \textit{normal} and
\textit{inverted} band structures of CdTe and HgTe are illustrated in
Fig.~\ref{fig2}(a) and Fig.~\ref{fig2}(b) respectively.\\

In this work, electronic properties of the $\left\langle001\right\rangle$ CdTe/HgTe/CdTe heterostructure are
calculated within an 8-band k.p framework that includes a linear coupling
between conduction and valence bands.~\cite{maciejko2011,sengupta2011} In the
calculations presented, the $z$-axis is normal to the heterostructure and is
also the confinement direction. The valence band edge $E_{v}$, Luttinger
parameters, and other related material properties are collected in
Table~\ref{table1}.~\cite{novik} The boundary conditions are imposed by
setting the wave function to zero at the edge of the device. 
Strain is added to the electronic Hamiltonian using deformation potentials defined in the
Bir-Pikus method.~\cite{schulman1986hgte,wu1985strain}

\begin{table}[h]
\caption{8-band k.p parameters for CdTe and HgTe. $E_{v}$, $E_{g}$, $P_{cv}$,
and V$_{so}$ are in units of eV. The remaining Luttinger parameters are
dimensionless constants and the effective mass is in units of the free
electron mass. } \
\begin{tabular}
[c]{|c|c|c|c|c|c|c|c|c|}\hline
Material & $E_{v}$ & $\gamma_{1}$ & $\gamma_{2}$ & $\gamma_{3}$ & $m^{*}$ &
$E_{g}$ & $P_{cv}$ & V$_{so}$\\\hline
CdTe & -0.27 & 5.372 & 1.671 & 1.981 & 0.11 & 1.606 & 18.8 & 0.91\\\hline
HgTe & 0.0 & -16.08 & -10.6 & -8.8 & -0.031 & -0.303 & 18.8 & 1.08\\\hline
\end{tabular}
\label{table1}
\end{table}

\section{Results and Discussion\label{sec:results}}

\begin{figure}[ptbh]
\includegraphics[scale=0.85]{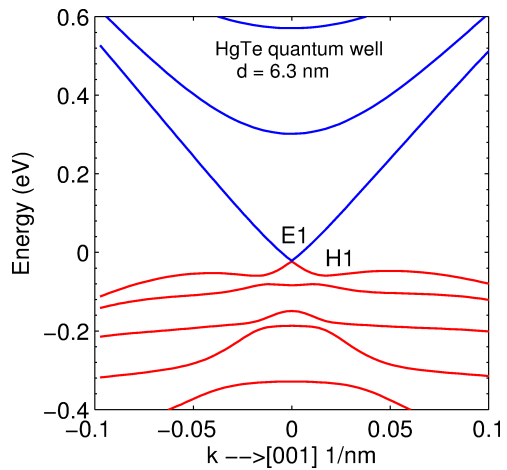}
\caption{Band structure of HgTe quantum
well of thickness $6.3~\mathrm{nm}$. At this width, the lowest conduction band (E1) and
highest valence band (H1) at the $\Gamma$ point are equal.}
\label{fig4}
\end{figure}

\begin{figure}[ptbh]
\includegraphics[scale=1.0]{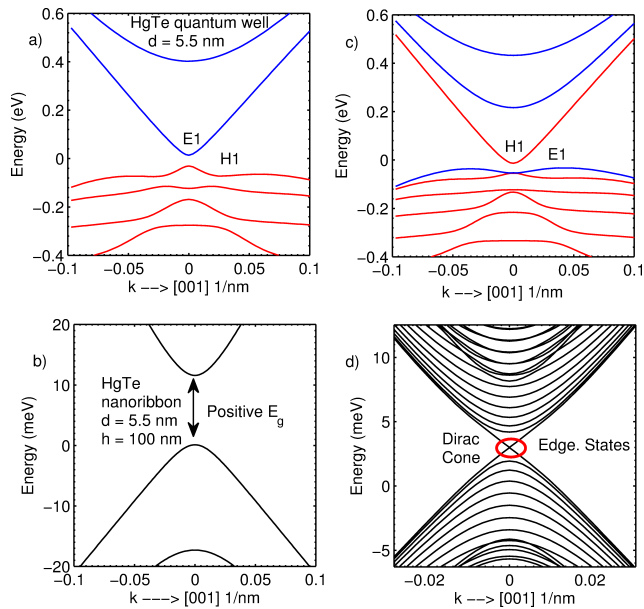}
\caption{Bandstructure of a HgTe quantum well of thickness $5.5~\mathrm{nm}$(a). A HgTe nano-ribbon formed out of this quantum well of thickness
$5.5~\mathrm{nm}$ and height of $100~\mathrm{nm}$ shows a positive band gap. Fig.~\ref{fig3}c shows the bandstructure of an inverted quantum well of thickness $10.0~\mathrm{nm}$. The corresponding quantum wire has a linearly dispersing (Dirac-cone) edge states (d).}
\label{fig3}
\end{figure}

\textbf{\emph{Comparison with experiment: \textbf{band gap and critical width}}}Experiments report that a CdTe/HgTe/CdTe quantum well heterostructure with a well width under $6.3~\mathrm{nm}$ exhibits a normal band order with positive $E_{g}$.~\cite{lu2011,brune2011}. The calculation of the present work confirms that the conduction states at $\Gamma$ are indeed located above the valence states and the energy gap is positive (Fig.~\ref{fig3}(a)). All band structure parameters used to reproduce the experimental observation were valid at 0 K. When the well width is exactly $6.3~\mathrm{nm}$, a Dirac system ~\cite{königarxiv} is formed in the volume of the device (Fig.~\ref{fig4}). \\

Beyond this \textit{critical well width} of $6.3~\mathrm{nm}$, the heterostructure has its bands fully inverted. The band profile has a reverse ordering of the \textit{s}-type and \textit{p}-type orbitals (Fig.~\ref{fig3}(c)) and $E_{g}<0$.
  
Accordingly, a nano-ribbon of width $100.0~\mathrm{nm}$ formed by quantizing the quantum well in its in-plane direction has a positive band gap (as shown in Fig.~\ref{fig3}b). Similarly, a nanoribbon of width $100.0~\mathrm{nm}$ constructed out of an inverted quantum well possesses gap-less TI edge states. The band structure of this situation is illustrated in Fig.~\ref{fig3}(d).\\
 
\begin{figure}[ptbh]
\includegraphics[scale=0.85]{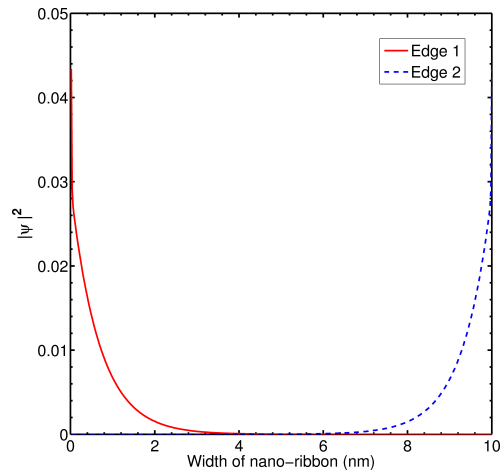}
\caption{Absolute value of the wave
functions $|\psi|^{2}$ of the two edge-states of Fig.~\ref{fig3}d.}
\label{fig6}
\end{figure}

The corresponding absolute value of the squared edge-state wave functions
is plotted in Fig.~\ref{fig6}. The absolute value of the wave functions for the two edge states is maximum at the edge and gradually decay in to the bulk. This establishes that they belong exclusively to the edge states. In conclusion, the band-gap closing Dirac cone shown in Fig.~\ref{fig4}
marks the transition from a positive band-gap to a negative one.

\textbf{\emph{Band nature at finite momenta:}} The inversion of bands in the
volume of the well is necessary for edge states with topological insulator
behavior. It is important to note however, that the process of inversion
happens only at the $\Gamma$ point. In the inverted dispersion plot
(Fig.~\ref{fig3}c), for momenta different from the $\Gamma$ point, the band
labeled with "H1" progresses from \textit{p} to \textit{s}-type. Similarly the
band labeled with "E1" changes character from \textit{s} to \textit{p}. Both
the bands, at a finite momentum acquire atomic orbital characteristics
associated with a normally ordered set of bands. TI behavior is therefore
restricted to a special set of momentum points where the band structure is
inverted. These set of points are collectively called the
time-reversal-invariant-momentum (TRIM) points.~\cite{teo2008}


\begin{figure}[!ht]
\centering
\includegraphics[scale=1]{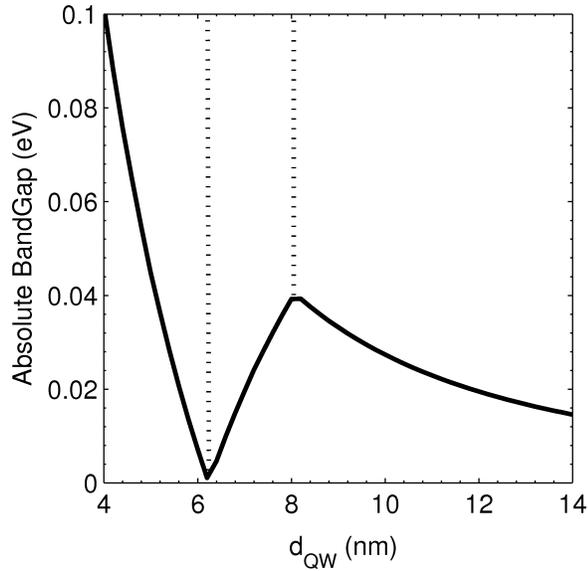}
\caption{Absolute value of the band gap
of a CdTe/HgTe/CdTe quantum well as a function of the well width. Well widths
larger than $6.3~\mathrm{nm}$ produce inverted band structures and can be exploited for
topological insulator devices. }
\label{fig7}
\end{figure}

\textbf{\emph{Well thickness continuously tunes the TI properties:}} With
increasing well width, the band gap decreases continuously until the HgTe well
thickness reaches $6.3~\mathrm{nm}$ (see Fig.~\ref{fig7}). This is due to the diminishing
confinement of the well's \textit{s} and \textit{p}-type bands. For well
thicker than $6.3~\mathrm{nm}$, the confinement is small enough such that the inverted
band ordering of HgTe is restored and the absolute value of the negative band
gap is increasing (see Fig.~\ref{fig7}).

At a HgTe well width of $8.2~\mathrm{nm}$, the
\textit{s}-type band drops even below the second confined \textit{p}-type
state. This re-ordering of bands with well thickness is summarized in
Table~\ref{table3} and illustrated in Fig.~\ref{fig7}.

\begin{table}[!h]
\caption{Orbital character of the top most valence band and lowest conduction
band in CdTe-HgTe-CdTe heterostructure depending on the well width d$_{QW}$.
The critical well width d$_{c}$ is the equal to $6.3~\mathrm{nm}$.}
\begin{tabular}
[c]{|c|c|c|}\hline
HgTe Well thickness & Highest Val.Band & Lowest Cond.Band\\\hline
d$_{QW}$ $<$ d$_{C}$ & p-type & s-type\\\hline
8.2 nm $>$ d$_{QW}$ $>$ d$_{C}$ & s-type & p-type\\\hline
d$_{QW}$ $>$ 8.2nm & p-type & p-type\\\hline
\end{tabular}
\label{table3}
\end{table}

\subsection{Stoichiometric and Temperature Control of Critical Width}

\begin{figure}[ptb]
\includegraphics[scale=0.85]{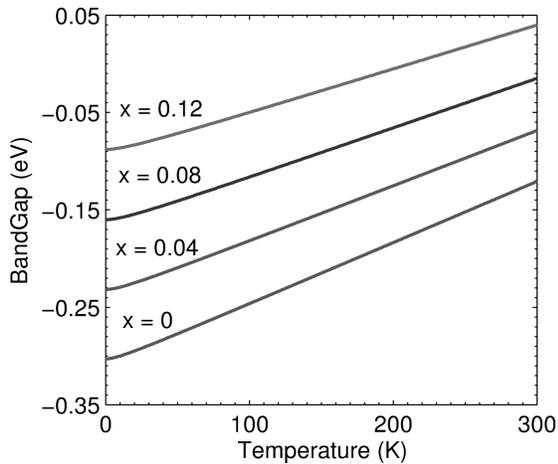}
\caption{Calculated band gap of bulk
Cd$_{x}$Hg$_{1-x}$Te as a function of stoichiometry and temperature. At
\textit{x}=0, the bulk band gap of HgTe ($-0.303~\mathrm{eV}$) is reproduced.}
\label{fig8}
\end{figure}

In the previous sections, it has been shown that the effective band gap of the
CdTe/HgTe/CdTe quantum well depends on the confinement and consequently on the
band gap difference of the well and barrier materials. Both, alloying and
temperature are known to influence the effective band gap. The band gap of
Cd$_{x}$Hg$_{1-x}$Te as a function of temperature~\cite{krishnamurthy1995temperature} $T$ and stoichiometry $x$~is
given by
\begin{equation}
E_{g}=-304+\frac{0.63T^{2}}{11+T}(1-2x)+1858x+54x^{2}.\label{egvt}%
\end{equation}
A plot for the band-gap variation for the Cd$_{x}$Hg$_{1-x}$Te alloy is given
in Fig.~\ref{fig8}. 

When the quantum well material of the original
CdTe/HgTe/CdTe structure is substituted by Cd$_{x}$Hg$_{1-x}$Te alloy, the
critical width becomes temperature and $x$ dependent. This is shown in
Fig.~\ref{fig6}~(a). Remarkably, all critical widths are equal or larger than
the intrinsic critical width of $6.3~\mathrm{nm}$. Higher concentration of CdTe in the
quantum well reduces the Darwin contribution from HgTe. Therefore, the band
inversion requires a wider HgTe region.

Alternatively, replacing the barrier material with Cd$_{x}$Hg$_{1-x}$Te also
allows tuning the confinement and consequently the critical width. It is shown
in Fig.~\ref{fig8}(b) that this replacement yields critical widths
smaller then the intrinsic $6.3~\mathrm{nm}$ if the temperature is allowed to attain values below 100 K.
For a Cd molar concentration of $x=0.68$ and $T=0~\mathrm{K}$,
the critical width dropped to $4.4~\mathrm{nm}$. This is due to the enhanced
Darwin contribution to the electronic properties with increased Hg content.
\begin{figure}
\includegraphics[scale=0.85]{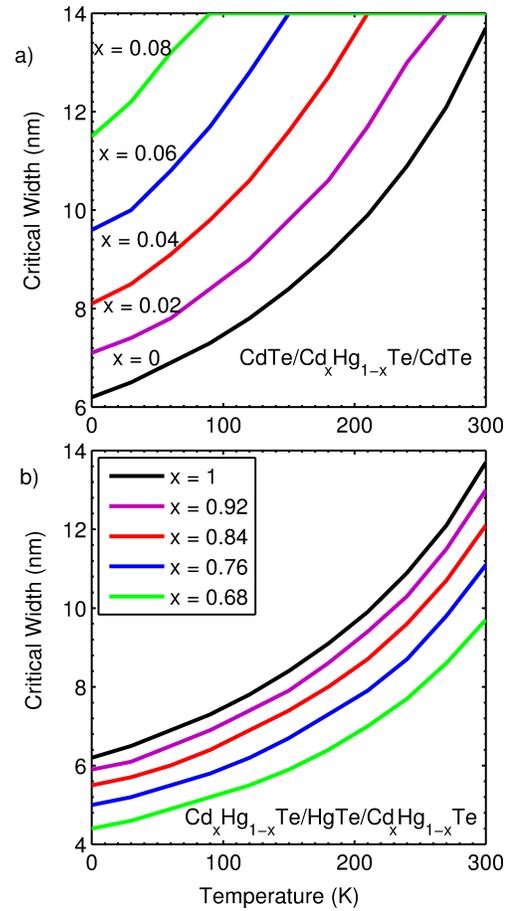}
\caption{Critical widths to get inverted
band structures of CdTe/Cd$_{1-x}$Hg$_{x}$Te/CdTe quantum wells (a) and
Cd$_{x}$Hg$_{1-x}$Te/HgTe/Cd$_{x}$Hg$_{1-x}$Te quantum wells (b) as a function
of temperature and stoichiometry $x$.}
\label{fig9}
\end{figure}

\subsection{Critical widths under different growth conditions}
\begin{figure}[h]
\includegraphics[scale=1]{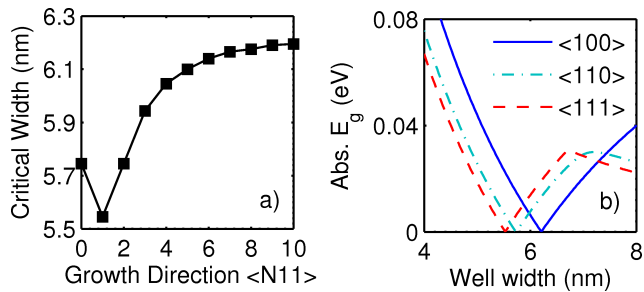} 
\caption{Critical widths of CdTe/HgTe/CdTe heterostructures grown along
$\left\langle N11\right\rangle $ direction as a function of $N$ (a). The bandgap closing for 
$\left\langle 100\right\rangle $, $\left\langle 110\right\rangle $, and $\left\langle 111\right\rangle $ grown CdTe/HgTe/CdTe at different well widths is shown in (b). Band gap closing at different well dimensions give the corresponding critical width.}
\label{fig10}
\end{figure}

Apart from alloy stoichiometry, the confinement also depends on the well and
barrier masses. A way to tune these effective confinement masses is by growing
the quantum well in different directions. The different masses then give
different effective well confinement and accordingly different critical
widths. This dependence is illustrated in Fig.~\ref{fig10}. It shows the critical width of
the CdTe/HgTe/CdTe quantum well in a sequence of growth directions.
The critical widths of the $\left\langle 111\right\rangle $ and $\left\langle110\right\rangle$
growth directions are $5.52~\mathrm{nm}$ and $5.72~\mathrm{nm}$ respectively. Both these values are smaller than the $\left\langle 001\right\rangle $ grown quantum well critical width of $6.3~\mathrm{nm}$.\\
	
\begin{figure}[!h]
\includegraphics[scale=0.85]{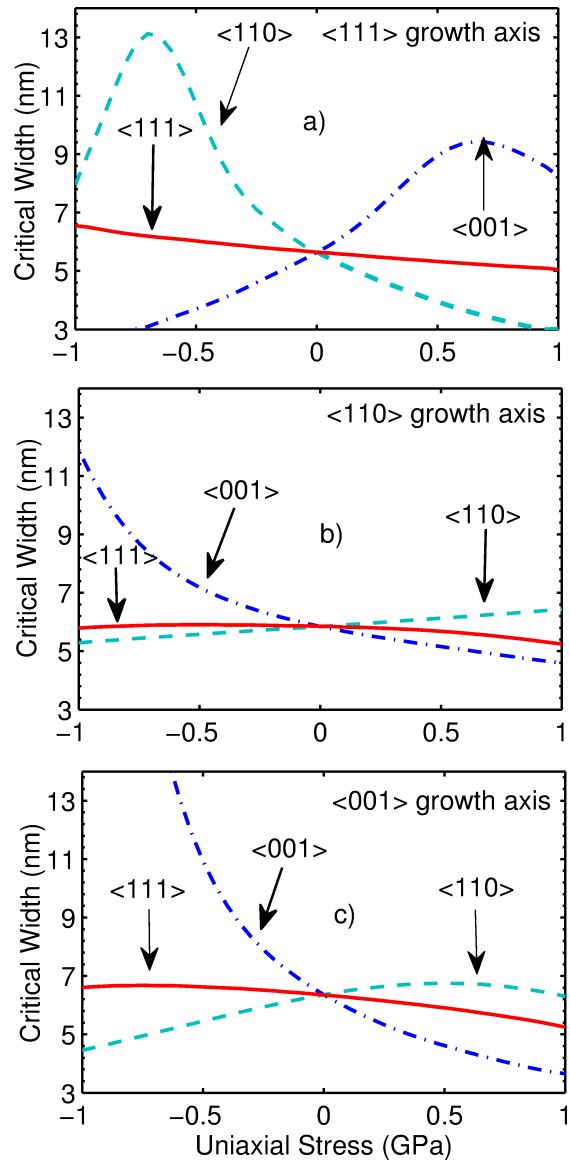} 
\caption{Critical widths of CdTe/HgTe/CdTe heterostructures grown along
$\left\langle 111\right\rangle $ (a), $\left\langle 110\right\rangle$ (b),
and $\left\langle 001\right\rangle$ (c) direction with uniaxial stress
applied along $\left\langle 111\right\rangle$ (solid), $\left\langle
110\right\rangle $ (dashed) and $\left\langle001\right\rangle$
\ (dash-dotted) direction. Key observations are summarized in
Table~\ref{table4a} and table~\ref{table4b}.}
\label{fig11}
\end{figure}

Alternatively, uniaxial stress can also tune the effective confinement masses.
As representative cases, CdTe/HgTe/CdTe quantum wells were grown along
$\left\langle001\right\rangle$, $\left\langle110\right\rangle$, and $\left\langle111\right\rangle$ directions. Each quantum well was then
subjected to uniaxial stress along $\left\langle001\right\rangle$, $\left\langle110\right\rangle$, and $\left\langle111\right\rangle$
directions. Uniaxial stress along $\left\langle001\right\rangle$, $\left\langle110\right\rangle$, and $\left\langle111\right\rangle$ was employed on
three sets of CdTe/HgTe/CdTe quantum wells grown along $\left\langle001\right\rangle$, $\left\langle110\right\rangle$,
and $\left\langle111\right\rangle$. The behaviour of the critical width for each case is shown in
Fig.~\ref{fig11}. The ideal stress orientation for each growth direction is
summarized in Tables~\ref{table4a} and \ref{table4b}.

\setlength{\tabcolsep}{10pt}
\begin{table}
\caption{The optimal tensile stress and growth conditions for CdTe/HgTe/CdTe
quantum wells to achieve the least (L), highest (H) and intermediate (I)
critical width, respectively.}
\begin{tabular}
[c]{|c|c|c|c|}\hline
& \multicolumn{3}{c|}{Tensile uniaxial stress} \\ \hline
Growth Axis & $\left\langle001\right\rangle$ & $\left\langle110\right\rangle$ & $\left\langle111\right\rangle$ \\\hline
$\left\langle001\right\rangle$ & L & H & I \\\hline
$\left\langle110\right\rangle$ & L & H & I \\\hline
$\left\langle111\right\rangle$ & H & L & I \\\hline
\end{tabular}
\label{table4a}
\end{table}

\setlength{\tabcolsep}{10pt} 
\begin{table}
\caption{The same list of conditions as in Table~\ref{table4a} but under
compressive stress.}
\begin{tabular}
[c]{|c|c|c|c|}\hline
& \multicolumn{3}{c|}{Compressive uniaxial stress}  \\[0.5ex]\hline
Growth Axis & $\left\langle001\right\rangle$ & $\left\langle110\right\rangle$ & $\left\langle111\right\rangle$ \\\hline
$\left\langle001\right\rangle$ & H & L & I  \\\hline
$\left\langle110\right\rangle$ & H & L & I  \\\hline
$\left\langle111\right\rangle$ & L & H & I  \\\hline
\end{tabular}
\label{table4b}
\end{table}

\subsection{Application of an external electric field}
\begin{figure}[!h]
\includegraphics[scale=1]{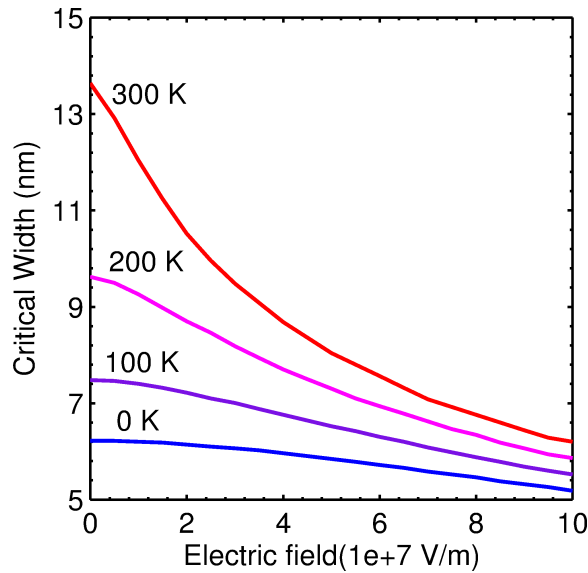}
\caption{Critical width for CdTe/HgTe/CdTe
quantum wells with varying strength of external electric fields in growth direction.}
\label{fig12}
\end{figure}

The application of an external electric field changes the confinement and the
band properties of the well states. In particular, the Rashba (structural
inversion asymmetry) effect gets enhanced by electric fields in growth
direction.~\cite{zhang2001rashba} Figure~\ref{fig12} shows the critical width
as function of the external electric field applied in the growth direction. The
critical width decays with increasing field for all considered temperatures.
The Rashba effect that supports the band inversion of HgTe gets increased by
the electric field. Consequently, smaller well widths are required to invert
the CdTe/HgTe/CdTe quantum well band structure when external electric fields
are present.

Since external electric fields can tune the critical width, the
concept of a TI-switch is obvious: A CdTe/HgTe/CdTe quantum well with a well
width that is close, but below the critical width can be switched by electric
fields between normal and inverted band order. Such a switching band structure
is expected to yield significant changes in the surface conductance, due to
the unique transport properties of topological insulator states.
 
\begin{figure}[h]
\includegraphics[scale=1]{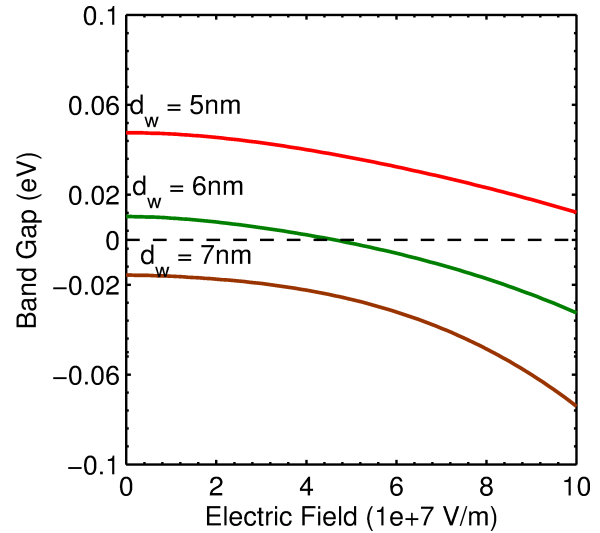}
\caption{Effective band gap of
CdTe/HgTe/CdTe quantum wells of different well thicknesses as a function of
applied electric field in growth direction. The dashed line depicts the
delimiter between normal and inverted band structures.}
\label{fig13}
\end{figure}

A first prototype of such a switch can be observed in Fig.~\ref{fig13} which shows
effective band gaps of CdTe/HgTe/CdTe quantum wells for various well
thicknesses under externally applied electric fields. Within the plotted range
of electric field magnitude, the CdTe/HgTe/CdTe quantum well with a width of
$6.0~\mathrm{nm}$ switches between normal and inverted band ordering. It is
worth mentioning that this switching behavior can be observed in
CdTe/HgTe/CdTe quantum wells grown in $\left\langle 001\right\rangle $ and
$\left\langle 111\right\rangle$ direction.

Such topological insulator based devices under an external electric field can be employed to act as a circuit element in a fast digital environment. When the bandgap is closed and TI properties are turned on, a high Fermi velocity for the carriers, (which is an essential attribute of TIs) on surface is able to transmit an electric signal faster than a conventional inter-connect. A seamless transition from a topological insulator to normal insulator using an external electric field as demonstrated above and shown in Fig.~\ref{fig13} enables it to forbid an easy passage of charge/electric signal. A normal insulator with a finite band gap will behave as an open circuit element.\\ 

\section{Conclusion\label{sec:conc}}
The present work investigates the conditions under which band inversion can
occur in a CdTe/HgTe/CdTe quantum well heterostructure. It is shown that this
band inversion is essential for topological insulator properties. In agreement
with experimental results, it is found that the HgTe quantum well has to be
thicker than $6.3~\mathrm{nm}$ to exhibit topological insulator properties. It
is examined in detail how the critical width depends on various device
parameters such as the growth direction, alloy stoichiometry, temperature,
uniaxial stress, and external electric fields. In particular the external
fields allow to switch the topological insulator properties of $\left\langle
001\right\rangle$ grown CdTe/HgTe/CdTe quantum wells. This result proposes a
new class of switching devices.


\begin{acknowledgements}
Computational resources from nanoHUB.org and support by National Science Foundation (NSF) (Grant Nos. EEC-0228390, OCI-0749140) are acknowledged. This work was also supported by the Semiconductor Research Corporation's (SRC) Nanoelectronics Research Initiative and National Institute of Standards \& Technology through the Midwest Institute for Nanoelectronics Discovery (MIND), SRC Task 2141, and Intel Corporation.
\end{acknowledgements}

\end{document}